\documentclass[aps,prd,twocolumn,groupedaddress,nofootinbib]{revtex4}  % for review and submission
\usepackage{graphicx}  % needed for figures
\usepackage{dcolumn}   % needed for some tables
\usepackage{bm}        % for math
\usepackage{amssymb,amsfonts,amsmath,physics}   % for math
\usepackage{float}
\usepackage{ulem}
\usepackage{tikz-feynman}
\tikzfeynmanset{compat=1.1.0}
\usetikzlibrary{arrows}
\tikzset{
    vertex/.style={circle,draw, minimum size=1.5em},
    edge/.style={->,> = latex'}
}

\usepackage[bookmarks, breaklinks, colorlinks,urlcolor=magenta, citecolor=red, 
linkcolor=blue]{hyperref}

\def\n {\nonumber}

\def\to {\rightarrow}
\def\sb{\color{blue}}

\newcommand{\bmt}{\begin{pmatrix}}
\newcommand{\emt}{\end{pmatrix}}
\newcommand{\ba}{\begin{array}{c}}
\newcommand{\ea}{\end{array}}
\newcommand{\be}{\begin{equation}}
\newcommand{\ee}{\end{equation}}
\newcommand{\bea}{\begin{eqnarray}}
\newcommand{\eea}{\end{eqnarray}}

\newcommand{\bi}{\begin{itemize}}
\newcommand{\ei}{\end{itemize}}

\newcommand{\baz}{\begin{array}{cc}}
\newcommand{\mathsym}[1]{{}}

\newcommand{\bt}{\begin{tabular}}
\newcommand{\et}{\end{tabular}}

\newcommand{\benu}{\begin{enumerate}}
\newcommand{\eenu}{\end{enumerate}}
\newcommand{\bee}{\begin{eqnarray}}
\newcommand{\eee}{\end{eqnarray}}
\newcommand{\beeq}{\begin{equation}}
\newcommand{\eeq}{\end{equation}}

\usepackage{xspace}

\newcommand{\bd}{\begin{displaymath}}
\newcommand{\ed}{\end{displaymath}}
\newcommand{\besub}{\begin{subequations}}
\newcommand{\eesub}{\end{subequations}}
\def\n{\nu}

\def\q2 {q^2}

\def\sb {\tilde{b} }

\def\bt{\begin{table}}
\def\et{\end{table}}
\begin{document}

% the following line is for submission, including submission to the arXiv!!
%\hspace{5.2in} \mbox{Fermilab-Pub-04/xxx-E}

\title{Imprint of the seesaw mechanism on feebly interacting dark matter and the baryon asymmetry}

\author{Arghyajit Datta}
%\email{datta176121017@iitg.ac.in}
\affiliation{Department of Physics, Indian Institute of Technology Guwahati, Assam-781039, India}

\author{Rishav Roshan}
%\email{rishav.roshan@iitg.ac.in}
\affiliation{Department of Physics, Indian Institute of Technology Guwahati, Assam-781039, India}

\author{Arunansu Sil}
%\email{asil@iitg.ac.in}
\affiliation{Department of Physics, Indian Institute of Technology Guwahati, Assam-781039, India}

%\date{\today}

\begin{abstract}

We show that the type-I seesaw, responsible for generating the light neutrino mass, itself is capable of accommodating one of the three right handed neutrinos as a freeze-in type of dark matter (DM) where the required smallness of the associated coupling is connected to the lightness of the (smallest) active neutrino mass. It turns out that (a) the non-thermal production of DM having mass $\lesssim \mathcal{O}(1)$ MeV (via decays of $W, Z$ bosons and SM Higgs)  consistent with relic density as well as (b) its stability determine this smallest active neutrino mass uniquely $\sim {\mathcal{O}}(10^{-12})$ eV. On the other hand, study of flavor leptogenesis in this scenario (taking into account the 
latest neutrino data and Higgs vacuum stability issue) fixes the scale of two other right handed neutrinos. 
\end{abstract}

\maketitle

%\section{\label{sec:level1}First-level heading}
% sections are not used for PRL papers

%%%%%%%%%%%%%%%%%%%%%%%%%%%
%\section{Introduction}
%\label{sec:intro}
%%%%%%%%%%%%%%%%%%%%%%%%%%
Among the various unresolved issues of present day particle physics and cosmology, perhaps the most pressing ones are the origin of tiny neutrino mass\cite{Fukuda:1998mi,Ahmad:2002jz,Ahn:2002up}, nature of dark matter (DM)\cite{Julian:1967zz,Tegmark:2003ud} and observed matter-antimatter asymmetry of the universe \cite{Aghanim:2018eyx}. In order to address these issues one has to anyway go beyond the standard model (SM) of particle physics, hence it would be very pertinent to search for a single minimal framework that can accommodate all these three problems together. To start with, one notices that the type-I seesaw mechanism\cite{Minkowski:1977sc,GellMann:1980vs,Mohapatra:1979ia,Schechter:1980gr,Schechter:1981cv} of neutrino mass generation provides a very promising platform for this. 
In this mechanism, three additional heavy SM singlet right handed neutrinos (RHN) are added to the SM particle content. 
A handful of attempts has been made in identifying one of them as dark matter without including any further beyond the SM fields and symmetries.  For example in the original $\nu$MSM model\cite{Asaka:2005pn,Asaka:2005an}, the lightest RHN (say $N_1$) is shown to be the DM having mass $\sim {\mathcal{O}}$ (keV). While the production of DM proceeds via Dodelson-Widrow (DW) mechanism\cite{Dodelson:1993je} incorporating the effective active-sterile neutrino mixing $\theta_1$, the ARS mechanism\cite{Akhmedov:1998qx} takes care of the observed baryon asymmetry via coherent oscillations of heavy RHNs. It turns out that the DW mechanism cannot make up the entire DM relic density taking into account the existing recent constraints on $\theta_1$ \cite{Hui:1996fh,Gnedin:2001wg,Weinberg:2003eg, Dolgov:2000ew,Abazajian:2001vt,Watson:2006qb,Essig:2013goa}. However, a variant of this incorporating a resonant production of DM via Shi-Fuller mechanism\cite{Shi:1998km} can still be operative \cite{Shi:1998km,Canetti:2012vf}. Though it bypasses the constraint on the mixing angle $\theta_1$, the mechanism suffers from an un-natural level of degeneracy required between the two heavy RHNs $N_{2,3}$.
Most of the other constructions with aim of identifying the RHN sector serving as the origin of DM and baryon asymmetry require additional fields and/or enhanced symmetry\cite{Okada:2010wd,DiBari:2016guw}.

In this letter, we stick to the most minimal construction of type-I seesaw while identifying $N_1$ as the feebly interactive massive particle (FIMP)\cite{Hall:2009bx,Bernal:2017kxu} type of DM and rest of RHNs are mainly responsible for generating light neutrino mass and matter-antimatter asymmetry. Interestingly we find that a sufficient production of $N_1$ can be obtained from the decays of $W, Z$ and SM Higgs $h$ which are intricately related to the specific entries of neutrino Yukawa matrix, $Y_{\nu}$ and in turn depend on the lightest active neutrino mass ($m_1$). These  entries, being involved in generating the respective active-sterile mixing $\theta_1$ associated to $N_1$, also control  possible decays of $N_1$. It turns out that an interplay between the production and decays of $N_1$ (such that it remains stable over the cosmological time scale) fixes the allowed range of DM mass ($M_1$) consistent with the stringent limits on $\theta_1$. 

The importance of our work lies in the fact that it provides perhaps the most minimal platform in the literature to address neutrino mass, dark matter and matter-antimatter asymmetry where the small coupling usually required for a FIMP realization is connected to the  smallness of the lightest active neutrino mass $m_1$. Such a connection is presented here for the first time to the best of our knowledge. It indicates an upper limit on $m_1$ as $m_1 \lesssim \mathcal{O}(10^{-12})$ eV. Interestingly, we find the DM relic turns out to be effectively independent of the DM mass within its allowed range. This opens up the possibility that the scenario can be tested if the recent and future experiments can measure $m_1$. 
As masses of $N_{2,3}$ ($M_{2,3}$) are unconstrained at this stage, we find that imposing an additional constraint on $Y_{\nu}$ as Tr[$Y^{\dagger}_{\nu} Y_{\nu}$] $<$ $\mathcal{O}$(1) (justified later) restricts the production of $N_1$ from the decays of $N_{2,3}$. This ambiguity of fixing $M_{2,3}$ is resolved once we incorporate flavor leptogenesis\cite{Abada:2006fw,Nardi:2006fx,Blanchet:2006be,Dev:2017trv}.

We start with the conventional type-I seesaw Lagrangian (in charged lepton diagonal basis) involving 
SM lepton ($l_L$) and Higgs ($H$) doublets by 
\bea
-\mathcal{L}_{\rm{Int}}= {(Y_{\nu})}_{\alpha i}\bar{l}_{L_{\alpha}} \tilde{H}N_{i}+\frac{1}{2}M_{i}\overline{N^{c}_{i}}N_{i}
+h.c.,
\label{lag}
\eea
with $i =1,2,3$ and $\alpha = e,\mu,\tau$. We assume the RHN mass matrix $M$ as diagonal with hierarchical masses. As a result of the electroweak (EW) symmetry breaking, the light neutrino mass matrix is given by the seesaw formula, $m_{\nu} = -m_D M^{-1} m^T_D$ which is diagonalized by $U^{\dagger} m_{\nu} U^* =$diag ($m_1, m_2, m_3$)$\equiv m_{\nu}^d$, where $U$ is the PMNS matrix \cite{Esteban:2020cvm} and ${(m_D)}_{ij} = {(Y_{\nu})}_{ij} v/{\sqrt{2}}$, where $v$ = 246 GeV. 

To begin with, we consider mass of the DM to so that possibility of its production from decays of the SM gauge bosons (via active-sterile neutrino mixing) 
and Higgs (via neutrino Yukawa interaction) remains plausible. On the other hand, masses of the remaining RHNs are assumed to be above the EW scale. Considering the fact that decay of $N_1$ can even proceed via the relevant active-sterile neutrino mixing ${m_D}_{i1}/M_1 \equiv V_{i1}$, we propose the following structure of neutrino Yukawa matrix at the leading order 
\bea
Y_{\nu} = \left(~ 
\begin{array}{*{13}{c}}
\cline{1-1}
\multicolumn{1}{|c|}{0}  
\\
\multicolumn{1}{|c|}{0} 
\\
\multicolumn{1}{|c|}{0}
 \\
\cline{1-1}
\end{array}~
\begin{array}{*{13}{c}}
\cline{1-2}
\multicolumn{1}{|c}{y_{e2}}& \multicolumn{1}{c|}{y_{e3}}  
  \\
\multicolumn{1}{|c}{y_{\mu 2}}& \multicolumn{1}{c|}{y_{\mu 3}} 
   \\
\multicolumn{1}{|c}{y_{\tau 2}}& \multicolumn{1}{c|}{y_{\tau 3}}
   \\
\cline{1-2}
\end{array}~
\right).
\label{yukawa}
\eea 
As a result of the vanishing left block (L$_B$), $N_1$ remains completely decoupled and hence absolutely stable while $N_{2,3}$ along with the right block (R$_B$) entries of $Y_{\nu}$ generate light neutrino mass via seesaw. This also ensures that the lightest active neutrino mass $m_1$ becomes zero and a vanishing $V_{i1}$ results. The entries of $Y_{\nu}$ can be written using the Casas-Ibarra (CI) parametrization \cite{Casas:2001sr}: 
\begin{align}
		m_D= -i~U D_{\sqrt{m}} R^T D_{\sqrt{M}},
		\label{CI para}
	\end{align}
where $U$ is the PMNS \cite{Zyla:2020zbs} mixing matrix, $D_{m} ~(D_{M})$ is the diagonal active neutrino (RHN) mass matrix: $m_{\nu}^d$ ($M$) and $R$ is a complex orthogonal matrix chosen to be of the form, 
\begin{align}
		R=\left(
		\begin{array}{ccc}
		1 & 0 & 0 \\
		0 & \cos \theta_R & \sin \theta_R \\
		0 & -\sin \theta_R & \cos \theta_R \\
		\end{array}
		\right), \label{eq:R}
	\end{align}
where $\theta_R$ is a complex angle in general. We employ the best fitted values \cite{Fogli:2006fw} of mixing angles, CP phase as well as mass-square differences to define the $U$ and $D_{\sqrt{m}}$. 

Under such a situation,  $N_1$ being completely segregated cannot be produced by any interaction (except gravitational one perhaps). This problem can be circumvented by perturbing $m_D$, $i.e.$ introducing small but nonzero entries in L$_B = (\epsilon_1, \epsilon_2, \epsilon_3)^T$ with $\epsilon_{i=1,2,3} \ll 1$. The order of smallness will be determined from the relic satisfaction of DM as well as from the stability of $N_1$. Note that origin of these 
$\epsilon_i$ can be associated\footnote{Alternatively, a tiny $m_1$ can be considered as an artifact of very small $\epsilon_i$.}  to a tiny $m_1$ or an additional angle of rotation (say $\varphi$) over $R$ or including both. In this work, without any loss of generality, we would like to pursue our analysis with small $m_1$ as the same result can be obtained from the use of $\varphi$. In this case, following Eq.(\ref{CI para}), it is seen that $\epsilon_i$ turns out to be proportional to $\sqrt{m_1 M_1}$. The DM phenomenology is almost independent to R$_B$ of $Y_{\nu}$.

With such a scenario in mind, the active neutrinos and $N_{2,3}$ remain in thermal equilibrium with other SM fields while $N_1$ is expected to be in out-of-equilibrium (due to its small coupling proportional to $\epsilon_i$) in the early universe having negligible abundance. Later, once the temperature goes down, the DM is expected to be produced non-thermally from decay of some massive particle or via annihilations. In this simplest seesaw set-up, we find subsequent to the EW symmetry breaking, the DM $N_1$ can be produced from the following dominant decays: 
\bea
\nonumber
W^{\pm} \to N_1 \ell^{\pm}_i, & Z \to N_1 \nu_i,  h \to N_1 \nu_i; & N_{i \neq 1} \to N_1 h(Z).
\nonumber
\eea
The relevant parts of the Lagrangian responsible for the above decays via the active sterile mixing $V = m_D M^{-1}$ are followed from the gauge interactions  
\begin{align}
-\mathcal{L_G}&\subset \frac{g}{\sqrt{2}} W_{\mu}^+\sum_{i,j=1}^{3}\Big[\bar{N}^c_i (V^{\dagger})_{ij}\gamma^{\mu} P_L \ell_j\Big]+\frac{g}{2 C_{\theta_w}} Z_{\mu}\times\nonumber\\
&\sum_{i,j=1}^{3}
\Big[ \bar{\nu}_i (U^{\dagger}V)_{ij}\gamma^{\mu} P_L N^c_{j} + \bar{N}^c_i (V^{\dagger}V)_{ij}\gamma^{\mu} P_L N^c_{j}\Big] + h.c., \nonumber\\
\label{gauge_int}
\end{align}
and Yukawa interactions 
\begin{align}
-\mathcal{L_Y}&\subset \frac{\sqrt{2}}{v} h \sum_{i,j=1}^{3}\Big[ \bar{\nu}_{i}(U^{\dagger}V)_{ij}M_j N_{j}
+  \bar{N}^c_{i}(V^{\dagger}U)_{ij} m_j\nu^c_{j}\nonumber\\ &+
 \bar{N}^c_{i}(V^{\dagger}V)_{ij}M_j N_{j}\Big] + h.c.,\label{ym}
\end{align}
where $\nu_i$ are active neutrino mass eigenstates. Origin of the most relevant mixing $V_{i1} = {m_D}_{i1}/M_1$ ($\equiv \epsilon_i \frac{v}{\sqrt{2}M_1}$) is traced back to $\epsilon_i$ entries of $Y_{\nu}$. 

We then employ the coupled set of Boltzmann equations involving $N_1$ and $N_{2,3}$ separately to study the 
evolution of their abundance ($Y_{N_i}$) till the present time, as
\begin{align}
\frac{dY_{N_1}}{dz}= &\frac{2 M_{pl} z}{1.66 m_h^2} \frac{g_{\rho}^{1/2}}{g_s}    
\Big[ \sum_{i=2,3} \Big(Y_{N_i} \sum_{x=Z,W} \big\langle\Gamma_{N_i \to N_1x}\big\rangle \Big)\nonumber\\ 
+&  \sum_{x=Z,h} Y_x^{eq} \big\langle\Gamma_{x\to N_1 \nu}\big\rangle + Y_W^{eq} \big\langle\Gamma_{W^{\pm}\to N_1\ell^{\pm}}\big\rangle\Big],\label{1BE}\\
\frac{dY_{N_i}}{dz}= & -\frac{2 M_{pl} z}{1.66 m_h^2} \frac{g_{\rho}^{1/2}}{g_s} 
\Big[(Y_{N_i}-Y_{N_i}^{eq})\langle\Gamma^D\rangle +  Y_{N_i}\nonumber\\& \sum_{x=h,Z}\langle\Gamma_{N_i \to N_1 x}\rangle\Big], ~(i =2,3),
\end{align}
with $z = m_h/T$. Here $\Gamma^D= \Gamma(N_i \to l H)+ \Gamma(N_i \to \bar{l} \bar{H})= \frac{ M_i}{8 \pi v^2}(m_D^{\dagger} m_D)_{ii}$ and $\big\langle \Gamma_{A\to BC}\big\rangle$ represents the thermally averaged  decay width\cite{Biswas:2016bfo}. All relevant decay widths are obtained from Eqs.\ref{gauge_int}-\ref{ym}. Note that the annihilations producing $N_1$ are very much suppressed ($\sim \epsilon^4_i$) compared to decay ($\sim \epsilon_i^2$) and hence are not included. At this stage, we presume $N_1$ to be stable over the cosmological time scale which will be justified in a while. Back reactions involving $N_1$ are not included as $N_1$ number density is vanishingly small to start with and for the same reason, terms proportional to $Y_{N_1}$ are also dropped. Substituting the abundance $Y_{N_1}(z_{\infty})$ after freeze-in, the relic density is obtained from,
\bea
\Omega_{N_1}h^2 &=& 2.755\times 10^5 \bigg(\frac{M_1}{\text{MeV}}\bigg)Y_{N_1}(z_{\infty}).
\label{relic_exp}
\eea

\begin{figure}[]
	
	\includegraphics[height=5cm]{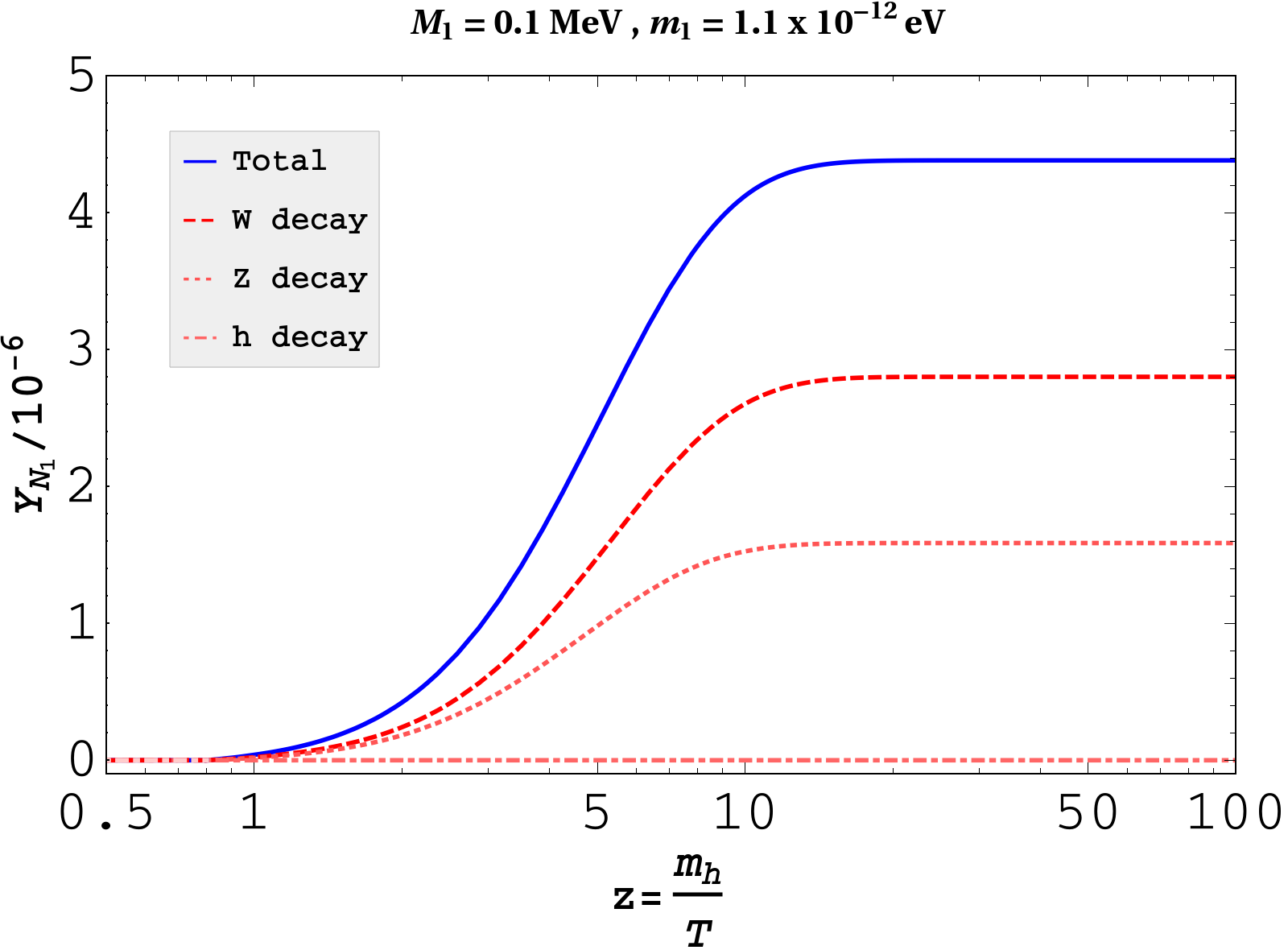}

	\caption{Abundance plot of $N_1$ with individual contributions (explained in inset) from different decays; final abundance (solid blue line) corresponds to the correct DM relic. }
	\label{fig:relic}
\end{figure}

The variation of the dark matter abundance $Y_{N_1}$ as function of $z$ is shown in 
Fig.\ref{fig:relic} where $Y_{N_1}$ (combined contribution as denoted by the solid blue line) reaches an asymptotic value, $Y_{N_1}(z_{\infty})$, so as to obtain the correct relic, $\Omega_{N_1} h^2 = 0.12$ \cite{Planck:2018vyg} via Eq. {\ref{relic_exp}}. Note that, we have parameters $m_1, M_{1,2,3}$ and $\theta_R$. 
In generating this plot, we have fixed $M_1$ at 0.1 MeV while $M_{2(3)}$ are kept at 3.5 (75) $\times 10^9$ GeV as deemed fit for generating correct baryon asymmetry via leptogenesis (discussed later). With such a choice of $M_1$,  $m_1 \sim 1.1 \times 10^{-12}$ eV is found to satisfy the relic implying $|\epsilon_i| \sim 10^{-15}$. We find that the production of $N_1$ is dominated by the decay of gauge bosons, in particular by $W^{\pm}$ decay\footnote{In \cite{Biswas:2016bfo, Borah:2020wyc, Coy:2021sse}, contributions of $W^{\pm}$ in $N_1$ production are estimated in the context of different extensions (gauge and/or fields) of the minimal set-up with SM and three RHNs.}, as emphasized in Fig.\ref{fig:relic}. The reason is the following. The production of $N_1$ from gauge and Higgs bosons depends on $\epsilon_i$ elements of $Y_{\nu}$ only (via $V$) whereas $N_1$ production\footnote{In a recent study \cite{Lucente:2021har}, it is shown that production of $N_1$ having mass in 1-80 keV range from $N_{2,3}$ can satisfy the relic.} from decay of $N_{2,3}$ involves a product of $\epsilon_i$ of L$_B$ and elements of R$_B$ (via $V^{\dagger} V$) as seen from Eqs. \ref{gauge_int}-\ref{ym}. While entries of L$_B$ are generated from $m_1$, elements of R$_B$ are controlled by the magnitude of $\theta_R$, mostly by Im$[\theta_R]$. We  find that any value of Im$[\theta_R]\lesssim$  5 keeps R$_B$ entries (or more precisely Tr[$Y_{\nu}^{\dagger} Y_{\nu}$]) below $\mathcal{O}(1)$.
We also notice that with larger Im$[\theta_R]$, entries of R$_B$ would increase significantly. Such a large $Y_{\nu}$ would be problematic not only from perturbativity but also due to the fact that EW vacuum becomes unstable \cite{Ghosh:2017fmr}.
Since DM production except from $N_{2,3}$ decays are anyway independent to entries of R$_B$, we refrain from quoting specific value for Re$[\theta_R]$ at this moment and reserve the related discussion for the leptogenesis part. 

As the DM $N_1$ mixes with the SM fields via the active-sterile mixing angle $V_{i1}$, we need to look for the all possible decay channels of it. There are three body decays \\
(a) [via off shell $W/Z$]: $N_1 \to l_1^- l_2^+ \nu_{l_2}$, $N_1 \to l^- q_1 {\bar q}_2$, $N_1\to l^- l^+ \nu_l$, $N_1\to \nu_l {\bar l}' l'$, $N_1 \to \nu_l q {\bar q}$, $N_1\to \nu_l {\nu}_{l'} {\bar \nu}_{l'}$, $N_1 \to \nu_l {\nu}_l {\bar \nu}_l$; \\
(b) [via off-shell Higgs]: $N_1 \to \nu_{\ell}  \bar{\ell} \ell$, \\
as well as (c) [radiative decay of $N_1$]: $N_1\rightarrow \gamma \nu$. \\ 
 Keeping in mind that the expected lifetime of $N_1$ must be greater than the age of the universe, it turns out that the most stringent constraint is obtained from (c), which can be translated on the active-sterile neutrino mixing $V_{i1}$ as \cite{Pal:1981rm,Barger:1995ty,Boyarsky:2009ix},
\bea
\theta_1^2&=&\sum_{i=1,2,3} |V_{i 1}|^2 \leq  2.8 \times 10^{-18} \bigg( \frac{\rm{MeV}}{M_1} \bigg)^5,
\label{theta1}
\eea

Below in Fig.\ref{fig:constraint}, we generate the relic contour plot in the $\theta^2_1-M_1$ plane drawn as the solid purple line, while the region in light blue is excluded from the above constraint. So at this point, we find $N_1$ as a successful FIMP type DM having mass below MeV. It is also interesting to note that the final DM relic density is independent to the mass of $N_1$. This observation stems from the fact that (a) the crucial parameter responsible for generating the dark matter abundance is $\epsilon_{i} \propto \sqrt{m_1 M_1}$ and (b) the dominant production of $N_1$ is from $W$ and $Z$ decays. The corresponding decay width (and hence $Y_{N_1}$ also) turns out to be proportional to $m_1/M_1$ (see Eq.(\ref{gauge_int})). Then the final relic density $\Omega_{N_1} h^2$ being related to $M_1 Y_{N_1}$, the $M_1$ dependence is cancelled out and $m_1$ is uniquely fixed to satisfy the relic. This leads to an interesting prediction for lightest active neutrino mass $\sim (1.07-1.12) \times 10^{-12}$ eV (considering the 3$\sigma$ range of DM relic density)  so that the model remains falsifiable in nature. Off course if one incorporates effect of both $m_1$ and the extra rotation $\varphi$, this value of $m_1$ serves as the upper limit of lightest neutrino mass.

We have verified that the non-thermality condition $\Gamma/H < 1$ at $T \sim m$ is satisfied where $\Gamma$ corresponds to the relevant decay width for a particular production channel of $N_1$ and $m$ is mass of the decaying particle.  Hence the DM particles produced (having mass range 1 keV - 1 MeV) cannot have sufficient energy to be associated with large free streaming length, thereby treated as cold dark matter, in contrary to the DW mechanism associated to warm dark matter ($\sim$ 2-10 keV). The lower limit on $M_1$ is considered as 1 keV to be in consistent with Tremaine–Gunn bound \cite{Tremaine:1979we} on sterile neutrino mass. A detailed study on the nature of DM in this range is beyond the scope of the letter. Finally, considering all these constraints, the range of DM mass turns out to be restricted within 1 keV-1 MeV.

\begin{figure}[h]

	\includegraphics[height=5cm]{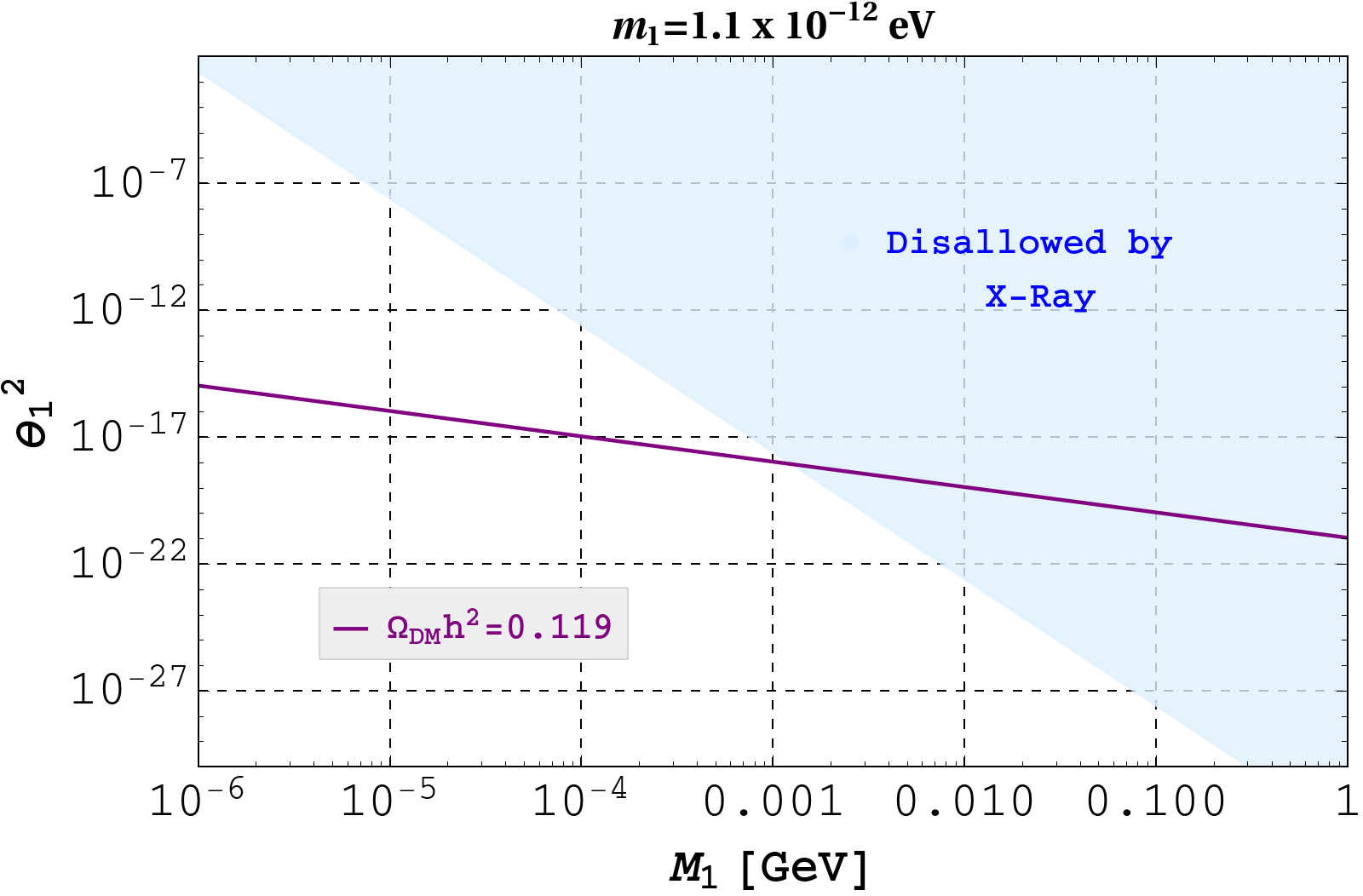}

	\caption{Relic satisfaction contour (solid purple line) in $\theta^2_1 - M_1$ plane. Constraint on $\theta_1^2$ from X-ray observation due to $N_1 \to \gamma \nu$ decay excludes the blue shaded region.}
	\label{fig:constraint}
\end{figure}

We now proceed to discuss the role of two other heavier RHNs, $N_{2,3}$ and their cosmological evolution. While they help in realizing the correct order of light neutrino mass and mixing, we find their contribution to DM production is almost negligible. That being said, their masses can be anywhere between a few hundred GeV to a very large scale. However, considering the fact that their decay can explain baryon asymmetry of the universe via leptogenesis, we can now have a complete picture including neutrino mass, dark matter and lepton asymmetry which will also tell us about these otherwise unspecified mass scales.  

Being heavier than the Higgs mass, $N_{2,3}$ are expected to decay into lepton doublet and Higgs via the Yukawa interaction of Eq.(\ref{lag}). This out of equilibrium decay along with the CP violation present in $Y_{\nu}$ will be crucial 
in leptogenesis. Note that Im[$\theta_R$] serving as the source of CP violation via Eq. (\ref{CI para}), apart for a subdominant contribution from Dirac CP phase in $U$, is the same one which mostly restricts the production of DM from the decay of $N_{2,3}$ while in tension with the EW vacuum stability. 

It is preferable to keep the heavy neutrino masses as low as possible in view of naturalness of hierarchy within RHNs, and hence we opt for flavor leptogenesis here. As $M_2 < M_3$, the CP asymmetry $\epsilon^{\rm{CP}}_{2\alpha}$ is effectively generated from the decay of $N_2$ to a specific flavor $l_{\alpha}$. Using the standard expression \cite{Nardi:2006fx}, we evaluate $\epsilon^{\rm{CP}}_{2\alpha}$ first and then 
\begin{figure}[h]
	\includegraphics[height=5cm]{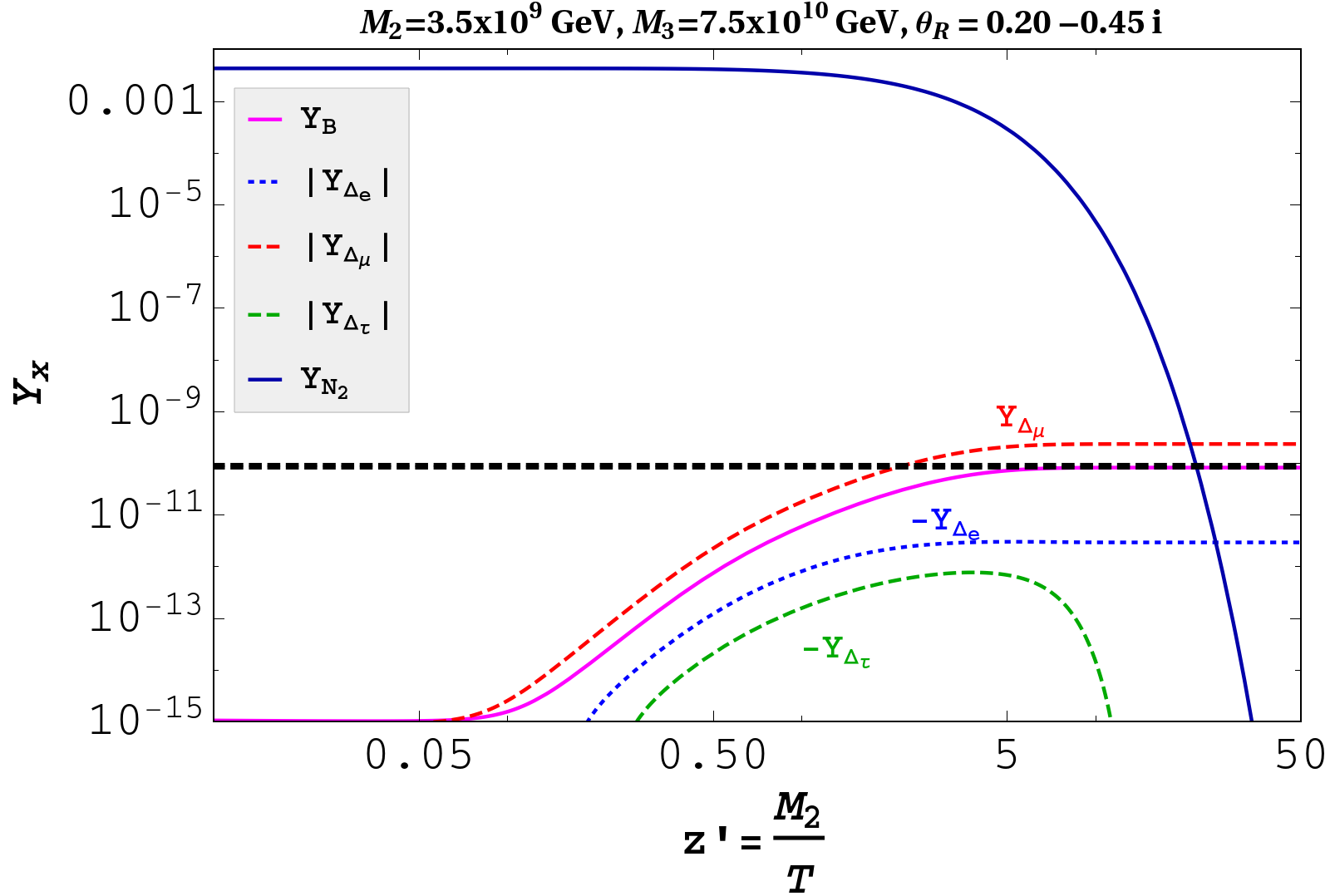}
	
	\caption{Evolution of individual flavor asymmetries as well as baryon asymmetry with respect to $z'=M_2/T$. Black dashed lines: range of observed baryon asymmetry of the universe.}
	\label{fig:lepto}
\end{figure}
proceed for estimating the final lepton asymmetry employing the set of Boltzmann equations 
\begin{widetext}
\begin{align}
s H z' \frac{d Y_{N_2}}{dz'}
=
-
\Big	\{	\left( \frac{Y_{N_2}}{Y_{N_2}^{\rm eq}}	-	1	\right)
(\gamma_{D} + 2 \gamma_{H_s}	+  4  \gamma_{H_t}&)\Big  \} \,,\label{be1}\\
s H z' \frac{d Y_{\Delta_{\alpha}}}{dz'}
=
-\Bigg\{
\left(	\frac{Y_{N_2}}{Y_{N_2}^{\rm eq}}	-1\right)
\epsilon_{2 \alpha}^{\rm{CP}} \gamma_{D}
+
K^0_{\alpha} \sum_{\beta} \Bigg[& \frac{1}{2}  (C^{\ell}_{\alpha \beta} +
C^H_{\beta}) \gamma_D 
+\left(\frac{Y_{N_2}}{Y_{N_2}^{\rm eq}}	-	1\right) \left(C^{\ell}_{\alpha \beta} \gamma_{H_s} + \frac{C^H_{\beta}}{2} \gamma_{H_t}\right) \notag\\
 &+\left(2 C^{\ell}_{\alpha \beta}+C^H_{\beta}\right) \left(\gamma_{H_t}+\frac{1}{2} \gamma_{H_s}\right)\Bigg]\frac{Y_{\Delta_{\beta}}}{Y^{\rm eq}}
\Bigg\}\label{eq:lep}
\end{align}
\end{widetext}
where $K^0_{\alpha}= \frac{(Y_{\nu}^*)_{\alpha 2} (Y_{\nu})_{\alpha 2}}{(Y_{\nu}^{\dagger} Y_{\nu})_{22}}$ is known as flavor projector and $C^{\ell}, C^H$ matrices connect the asymmetries in lepton and Higgs sectors to asymmetries 
in $\Delta_{\alpha} = B/3 - L_{\alpha}$ expressed in terms of $Y_{\Delta{\alpha = e, \mu, \tau}}$. Here $\gamma_X$ is the corresponding reaction rate density \cite{Plumacher:1996kc}. The final baryon asymmetry is obtained as  
$Y_B= (28/79) \sum_{\alpha} Y_{\Delta_{\alpha}}$.

Fig.\ref{fig:lepto} depicts the variation of individual components of lepton asymmetry $Y_{\Delta_{\alpha}}$ as well as the total baryon asymmetry  $Y_B$ with respect to $z' = M_2/T$. It turns out that the observed baryon asymmetry results for the lowest possible value of $M_2 = 3.5 (5) \times 10^9$ GeV with Re$[\theta_R]=0.2 (0)$ and Im[$\theta_R$] = -0.45 (-0.4). The corresponding value of $M_3$ is found to be 7.5 (15) $\times 10^{10}$ GeV. 
At this temperature range $\sim M_2$ value, muon and tau Yukawa interactions come to equilibrium and hence lepton asymmetries along all the flavor directions become relevant (see Fig.\ref{fig:lepto}). We also infer from Fig.\ref{fig:lepto} that the abundance of $Y_{N_2}$ with such a large $M_2$ is falling sharply as temperature decreases and hence is expected to be vanishingly small in the EW broken phase where $N_1$ production is mainly taking place from the decay of the SM gauge bosons. This along with the fact that production of $N_1$ from 
$N_2$ decay is also suppressed (via $V^{\dagger}V$ as stated earlier) eventually indicate that $N_2$ contributes effectively
nothing to $N_1$ production as seen from the right hand side of Eq. \ref{1BE} (first term). The same conclusion holds 
for $N_3$ as well.

In summary, we have shown that the conventional type-I seesaw scenario itself has the potential to offer a FIMP type of dark matter in the form of lightest RHN, the relic density of which is mainly governed by decay of the SM gauge bosons in the electroweak symmetry broken phase. With the hypothesis that in the limit of zero lightest active neutrino mass the dark matter is absolutely stable, implies that production and stability of the dark matter both are effectively controlled by the tiny active neutrino mass. The proposal predicts an upper bound on this lightest neutrino mass as $m_1 \lesssim {\mathcal{O}}$(10$^{-12}$) eV which makes it falsifiable if ongoing (or future) experiments such as KATRIN \cite{Aker:2021exx} and PROJECT-8 collaboration \cite{Esfahani:2017dmu} succeed to probe it. In this way, the smallness of couplings involved in a generic FIMP type model, related to dark matter production, can now be connected with the lightness of active neutrino mass $m_1$. While we find the DM mass $\sim$ 1 keV- 1 MeV satisfies the correct relic density as well as the stringent limits from X-ray observation, the dark matter phenomenology does not restrict the mass scales of two other heavy RHNs. Then we incorporate the flavor leptogenesis scenario to show that they can be $\sim 10^{9-10}$ GeV to explain the baryon asymmetry of the universe. So the minimal set-up of type-I seesaw can simultaneously address the origin of neutrino mass, non-thermal production of dark matter and matter-antimatter asymmetry without any additional fields. The presence of active-sterile neutrino mixing in the set-up is suggestive of the rare lepton flavor violating decays. The most relevant branching ratio in this context is related to $\mu \rightarrow e\gamma$ which turns out to be function of active-sterile neutrino mixing $V$ as well as RHN masses $M_i$. Employing values of $M_i$ (used in producing Fig.\ref{fig:relic} or Fig. \ref{fig:lepto}) and corresponding $V$ elements, the branching ratio of 
$\mu \rightarrow e\gamma$ is found to be negligibly small  compared to the present experimental limit \cite{MEG:2016leq}. We also evaluate the effective neutrino mass parameter involved in the half-life of neutrinoless double beta decay (function of $m_i$, lepton mixing angles and phases) and find it to be insignificant. 

\bibliography{ref.bib}

\end{document}